\documentclass{elsart1p}
\usepackage{graphicx}
\usepackage{amssymb}
\begin{document}
\begin{frontmatter}
%
% Title, authors and addresses
%
% use the thanksref command within \title, \author or \address for footnotes;
% use the corauthref command within \author for corresponding author
% footnotes;
% use the ead command for the email address,
% and the form \ead[url] for the home page:
% \title{Title\thanksref{label1}}
% \thanks[label1]{}
% \author{Name\corauthref{cor1}\thanksref{label2}}
% \ead{email address}
% \ead[url]{home page}
% \thanks[label2]{}
% \corauth[cor1]{}
% \address{Address\thanksref{label3}}
% \thanks[label3]{}
%
\title{Lattice gauge theory in technicolor}
%
% use optional labels to link authors explicitly to addresses:
% \author[label1,label2]{}
% \address[label1]{}
% \address[label2]{}
%
\author{Benjamin Svetitsky}
\address{Raymond and Beverly Sackler School of Physics and Astronomy, Tel Aviv University, 69978 Tel Aviv, Israel}
\begin{abstract}
% Text of abstract
The methods of lattice gauge theory may be applied to gauge theories besides QCD, in fact to any gauge group and any representation of matter fields (as long as the coupling is not chiral).  Such theories are useful for model building beyond the Standard Model, for instance in technicolor models.  We have carried out Monte Carlo simulations of the SU(3) gauge theory with color sextet fermions.  Our result for its discrete beta function indicate an infrared fixed point that makes the theory conformal rather than confining.  The lattice theory's phase diagram shows no separation of chiral and confinement scales, measured when the quark mass is nonzero.
\end{abstract}
\end{frontmatter}
%
% main text
\section{Introduction}

The past year has seen an acceleration in the application of lattice gauge theory to models other than QCD, particularly to field theories
that may find application in describing physics beyond the Standard Model
\cite{LGT}.
Gauge groups other than SU(3), and quark representations other than the
fundamental, figure often in BSM models.
We need to go beyond perturbation theory to verify whether a given model
is suitable for technicolor, for instance, or whether it fails to confine
quarks and thus becomes a theory of ``unparticles.''
Supersymmetry also cries out for nonperturbative treatment.
Current studies include QCD with various
numbers of flavors, the SU(2) gauge theory with fermions in the adjoint representation, and the supersymmetric
SU(2) gauge theory.
I will describe here our study of the SU(3) gauge theory with color sextet
quarks \cite{us1,us2}.

One should note significant limitations of lattice gauge theory or of its current
implementations.
For one thing, chiral gauge theories, so basic to weak interactions and beyond, have not yet
found a lattice formulation (except for the Abelian theory \cite{Luscher}).
Also, supersymmetric theories other than the simplest cannot be
studied on a lattice without breaking supersymmetry severely.

\section{SU(3) gauge theory with color sextet quarks}

We chose to study this model in order to settle two issues.
The first is the question of the existence of an infrared fixed point (IRFP)
that would render the infrared theory scale-invariant (rather than confining).
The second is the possibility of scale separation, where the large
Casimir operator of the sextet rep causes chiral symmetry
to break spontaneously at an energy scale higher than the confinement scale.
I will describe our work on the first question; for the second, suffice it to
say that we have found no evidence for scale separation, but the issue isn't
closed.

Caswell \cite{Caswell} noticed long ago that if you take QCD and increase the number of flavors
$N_f$ beyond 8 then the two-loop term in the $\beta$ function is positive.
This makes the $\beta$ function cross zero at a nonzero coupling $g_*$.
If one takes the two-loop $\beta$ function to be the whole story, then 
this theory has an IRFP.  The theory would be
asymptotically free at high energies and scale-invariant at low energies,
precluding confinement, chiral symmetry breakdown, a massive
particle spectrum, and, indeed, any particles at all.

The same happens if you take quarks in the sextet rep with $N_f=2$:
The two-loop $\beta$ function indicates an IRFP at $g_*^2\simeq10.4$.
This is a rather strong coupling, which makes a two-loop result suspect.
Nonetheless, the IRFP might survive in the full theory.

Or not:
As one comes down in energy, the strengthening coupling might cause
the chiral condensate to appear, whereupon the fermions would decouple
and their contribution to the $\beta$ function would disappear.  The theory would subsequently evolve as a pure gauge theory, consistent with confinement at
large distances and the chiral condensate.
This is what is needed for technicolor, and indeed this model has been
proposed as a theory of ``next-to-minimal'' walking technicolor \cite{FS}.

As it turns out, our numerical results argue for the first possibility,
that the IRFP is real and there is no confinement.

\section{Schr\"odinger functional method}

Since the $\beta$ function is at the heart of the matter, we calculate
it \cite{us1} (actually, a discrete version) on the lattice using the Schr\"odinger functional method \cite{SF},
which is an implementation of the background field method.
On a four-dimensional lattice of size $L^4$, we fix the gauge potential on the spacelike
boundaries at times $t=0$ and~$L$, which creates a classical background electric
field throughout.
We then calculate the free energy $\Gamma\equiv-\log Z$ (more precisely, 
a derivative of $\Gamma$ with respect to some parameter).
Comparing the free energy to the classical action $S^{cl}$
of the background field then yields the running coupling $g^2(L)$ at the
scale $L$, via $\Gamma=g(L)^{-2}S^{cl}$.
Comparison of two lattices, with size $L$ and $2L$, at the same lattice
spacing $a$,
gives the {\em discrete beta function\/} (DBF) for rescaling by a factor
of two,
\begin{equation}
B(u,2)=\frac{K}{g^2(2L)}-\frac{K}{g^2(L)},
\end{equation}
a function of $u\equiv K/g^2(L)$, where $K\simeq37$ is connected to
the classical action of our background field.

\begin{figure}
\begin{minipage}{.5\columnwidth}
\includegraphics*[width=.9\columnwidth]{SSF7.eps}
\caption{Discrete beta function for $4\to8$.\label{fig1}}
\end{minipage}
\begin{minipage}{.5\columnwidth}
\includegraphics*[width=.94\columnwidth]{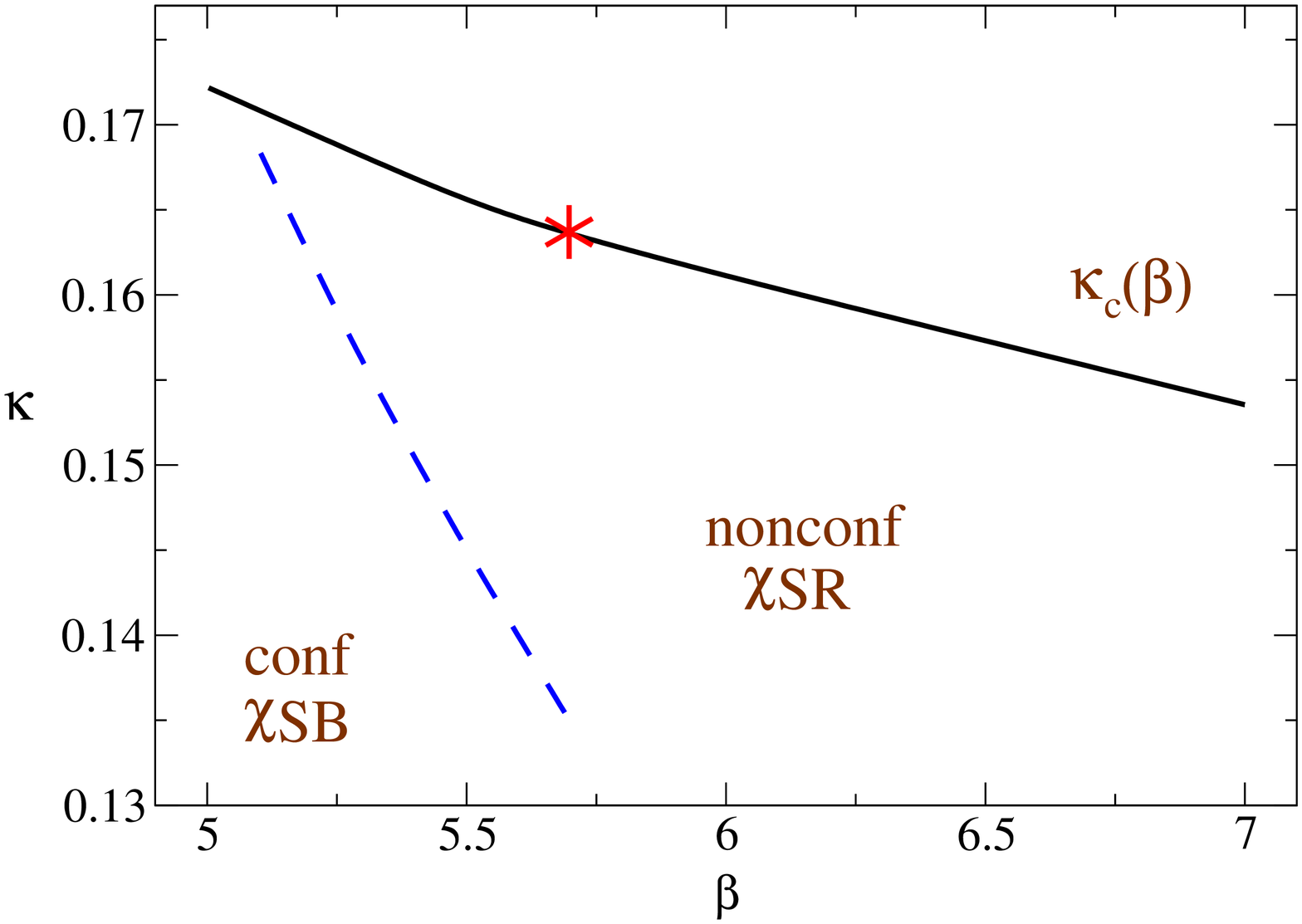}
\caption{Phase diagram with nonzero quark mass.
\label{fig2}}
\end{minipage}
\end{figure}

The result of comparing a lattice of size $4^4$ to one of size $8^4$ is shown
in Fig.~\ref{fig1}.  
The DBF crosses zero at $g^2\simeq2.0$, a far cry from
the two-loop estimate of $g^2\simeq10$ and, all in all, a fairly weak
coupling.
If confirmed on larger lattices, this will indicate that the
IRFP is a nonperturbative reality, and the theory is conformal (i.e.,
scale invariant) at large distances---{\em not\/} a candidate for
technicolor.

\section{Caveat}

Lattice studies {\em always\/} have to be pushed to larger lattices.  Here
this is equivalent (for given physical lattice size $L$) to ever smaller
lattice spacing.
A lattice with $4^4$ sites is rather small; one might find reassurance in the fact that we find the same result in comparing $6^4$ with $8^4$.  We are working
on the DBF for $6\to12$, which, together with the $4\to8$ result
shown, will furnish the beginning of a scaling
study for scale ratio 2.

Some more insight is offered \cite{us2}
by a look at the theory with nonzero quark
mass $m_q$.  Here we can look for finite-temperature phase transitions to
see what the physical scales are (see Fig.~\ref{fig2}).  We find
that (1) there is only one phase transition (dashed line) for any lattice size studied,
meaning no scale separation; and (2) our IRFP (the star on the $\kappa_c$ line,
along which $m_q=0$) is always in the
weak-coupling phase, indicating that a single-coupling $\beta$ function
may indeed give a valid description of the physics.
The properties of the conformal theory at the IRFP offer fascinating
questions for future work.

\end{document}